\documentclass{article}
\usepackage{spconf,amsmath,epsfig,url,color}

\title{Filtering Internal tides from wide-swath altimeter data using Convolutional Neural Networks}
\name{\begin{tabular}{c}Redouane Lguensat$^{*1}$, Ronan Fablet$^{2}$, Julien Le Sommer$^1$, Sammy Metref $^{1}$, Emmanuel Cosme$^{1}$, \\Kaouther Ouenniche$^{2}$, Lucas Drumetz$^{2}$, Jonathan Gula$^{3}$\end{tabular}\thanks{© 2020 IEEE.  Personal use of this material is permitted.  Permission from IEEE must be obtained for all other uses, in any current or future media, including reprinting/republishing this material for advertising or promotional purposes, creating new collective works, for resale or redistribution to servers or lists, or reuse of any copyrighted component of this work in other works.}}
\address{$^1$ Universit\'e Grenoble Alpes, CNRS, IRD, Grenoble INP, IGE; Grenoble, France\\$^2$ IMT Atlantique, LabSTICC, Universit\'e Bretagne Loire;  Brest, France \\ $^3$ Ifremer, LOPS; Brest, France
}
\begin{document}
\maketitle
\begin{abstract}
The upcoming Surface Water Ocean Topography (SWOT) satellite altimetry mission is expected to yield two-dimensional high-resolution measurements of Sea Surface Height (SSH), thus allowing for a better characterization of the mesoscale and submesoscale eddy field. However, to fulfill the promises of this mission, filtering the tidal component of the SSH measurements is necessary. This challenging problem is crucial since the posterior studies done by physical oceanographers using SWOT data will depend heavily on the selected filtering schemes. In this paper, we cast this problem into a supervised learning framework and propose the use of convolutional neural networks (ConvNets) to estimate fields free of internal tide signals. Numerical experiments based on an advanced North Atlantic simulation of the ocean circulation (eNATL60) show that our ConvNet considerably reduces the imprint of the internal waves in SSH data even in regions unseen by the neural network. We also investigate the relevance of considering additional data from other sea surface variables such as sea surface temperature (SST).
\end{abstract}
\section{Introduction}
\label{sec:intro}
This study is conducted within the framework of the next-generation Surface Water Ocean Topography (SWOT) satellite altimetry mission. The SWOT altimeter will rely on its wide-swath capacities to provide unprecedented two-dimensional maps of Sea Surface Height (SSH) down to a 10 km effective resolution. These measurements are expected to drastically improve the quality of SSH mapping and therefore enhance our understanding of the mesoscale and submesoscale dynamics of the upper ocean \cite{durand2010surface,fu2009swot,fu2008observing}. SWOT data will also yield valuable information on tidal components of the SSH signal such as shelf tides, coastal tides and open-ocean internal tides. Separating the tidal and non-tidal components of the SSH data is a critical issue for the physical oceanography community to ensure a proper exploitation of SWOT data for studying mesoscale and submesoscale flows \cite{arbic2015tides}.

At the best of our knowledge, Torres et al. \cite{torres19} have been the only to propose a method to extract Internal Gravity Waves (IGWs) from two-dimensional SSH snapshots. Their method is based on the identification of spectral slope discontinuities that separates IGWs at small scales from balanced motions at large scales. As such, it can recover the spectral content of the mesoscale and submesoscale flows only for spatial scales, where they strongly dominate over the IGWs, which may limit the use of this method. For instance, for this reason, only summer SSH data are investigated in \cite{torres19}.


\begin{figure}[t]
    \centering
    \includegraphics[scale=0.28]{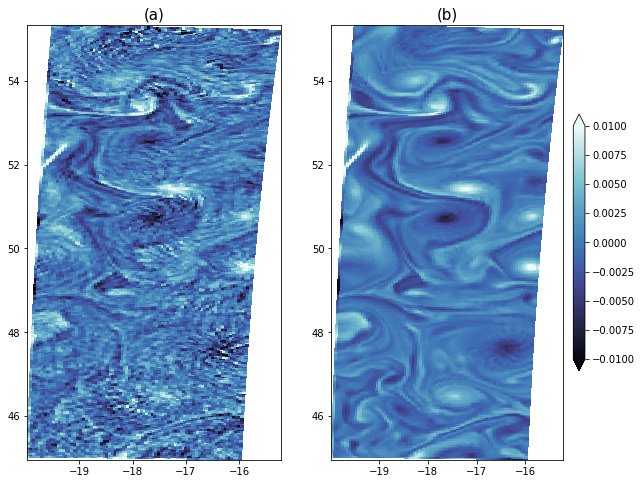}
    \caption{Footprints of internal tides on the Laplacian of a SSH snapshot: (a) Original data (b) 24h-averaged data}
    \label{fig:laplacian}
\end{figure}{}

Here, we state the filtering of tide signals in SSH fields as a supervised machine learning issue and explore deep learning techniques \cite{lecun2015deep}. We present a case-study, which encompasses both summertime and wintertime data, based on high resolution oceanic numerical simulation data \cite{madec2015nemo}.
Our main contributions are as follows: i) investigating the extent to which ConvNets can be relevant schemes for filtering IGWs, ii)
illustrating the relevance of the proposed ConvNets for both summertime and wintertime sea surface dynamics,  
iii) evaluating the potential gain of 
of considering multi-temporal data or multi-modal synergies. 

This paper is organized as follows. Section 2 presents the dataset considered as a testbed in this study. We describe the proposed ConvNet schemes in Section 3. Numerical experiments are reported and discussed in Section 4, and finally a conclusion is drawn in Section 5.

\begin{figure*}[t]
    \centering
    \includegraphics[scale=0.5]{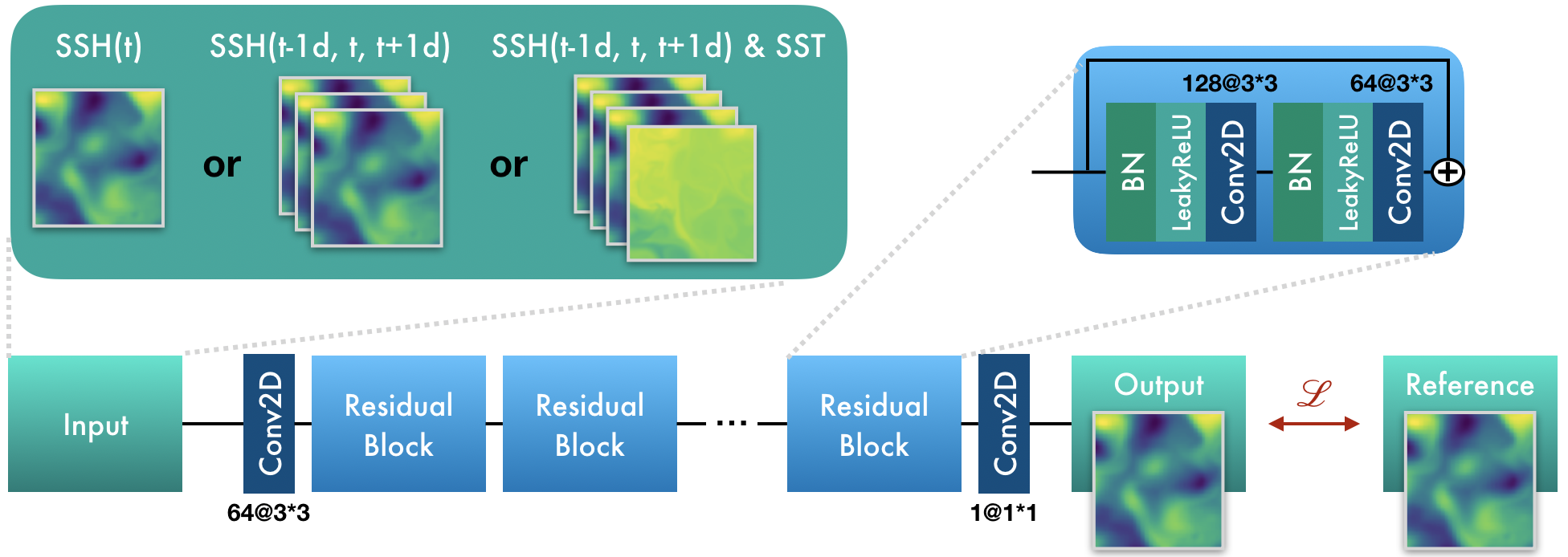}
    \caption{Illustration of the ConvNet based approach for filtering IGWs. A Residual block consists of two series of Batch Normalization (BN), LeakyReLU activation followed by a Conv layer. The notation M@H*H means M filters of size H*H.}
    \label{fig:convnet}
\end{figure*}{}

\section{Data}
\subsection{Data preparation}
Advancements in ocean numerical modeling have reached a state where realistic numerical simulation with Petabytes (10$^{15}$ bytes) of data are available and represent a real opportunity for machine learning based techniques. In this work, we consider the high-resolution eNATL60 (North Atlantic, hourly temporal resolution, 1/60$^\circ$ horizontal resolution) configuration of the NEMO (Nucleus for European Modelling of the Ocean) modelling system \cite{madec2015nemo}. Within the framework of the upcoming SWOT mission, 
we degrade the resolution by a factor of 3. We here focus on a subregion of the North Atlantic, namely, the OSMOSIS region (44.821$^\circ$N-55.363$^\circ$N, 20.016$^\circ$W-10.008$^\circ$W). The OSMOSIS region has a weaker large-scale
component compared to highly energetic regions such as the Gulf Stream \cite{buckingham2016seasonality}. This makes OSMOSIS a relevant region for assessing the ability of the investigated filtering schemes to recover small-scale dynamics which are the main component of interest awaited from SWOT.

We may point out that mesoscale and submesoscale upper ocean dynamics are known to be seasonally-dependent. Hereafter, we refer to "JAS" (July, August, September) for summer and "JFM" (January, February, March) for winter, when considering a season-related analysis.
The one-year span of eNATL60 resorts in 24$\times$ 90 = 2160 images for each dataset. We also split the region into several non-overlappling $64\times64$ patches. Splitting the dataset into train/val/test splits is done spatially which is arguably more challenging in our context than a temporal splitting, since oceanic patterns may strongly depend on the geographic area (different bathymethry, Rossby radius, etc.). We consider a spatial split according to the latitude. Overall, 5 boxes from 44.821$^\circ$N-52$^\circ$N were used to train our models resulting in 5*2160 =  10800 patches, and 1 box from the northwestern area of the OSMOSIS region is considered as the test region.

\subsection{Reference data used for supervised training}

Within a supervised training framework, we exploit the hourly time sampling of the eNATL60 simulation to build a tide-free reference using a 24-hour time averaging. This straightforward approach is commonly used in the analysis of the outputs of oceanic numerical models \cite{richman2012inferring}. More complex time filtering approaches could be considered to improve the generation of the tide-free samples.


An example of the Laplacian of a SSH snapshot from the considered dataset and its corresponding 24h filtered SSH is shown in Figure \ref{fig:laplacian}. we show Laplacian fields, which relate to the vorticity and clearly exhibit the footprints of the IGWs.   


\section{Methods}
\subsection{Convolutional Neural Networks}
Inspired by the connectivity patterns of neurons in animals' visual cortex, convolutional neural networks (hereinafter ConvNets) are one of the main and most important breakthroughs in neural networks literature \cite{lecun1995convolutional}. They have rapidly become a key component of state-of-the-art deep learning architectures in numerous computer vision tasks. 
Mathematically speaking, ConvNets in their basic form consist of a cascade of convolutional layers where the output of layer $k$ (consisting in so called feature maps) is a function of an affine transformation of the previous layer output: 
\begin{equation}
h_{i j}^{k}=f \left(\left(W^{k} * x\right)_{i j}+b_{k}\right),
\end{equation}
where $*$ denotes the convolution operation, $f$ is a nonlinear function (a common choice is the Rectifier Linear Unit ReLU$(x)=max(0,x)$). Weights $W^{k}$ and biases $b_{k}$ are the parameters of the ConvNets to be inferred. The learning step comes to solve for a minimization issue with regard to the parameters of the ConvNet, for instance using the Stochastic Gradient Descent (SGD) with the backpropagation algorithm. The interested reader can refer to \cite{lecun2015deep} and references therein for more reading about ConvNets and their mathematical properties. 

Here, we focus on ResNet, a specific class of ConvNets which rely on 
residual blocks \cite{he2016identity,he2016deep}. Let $\mathcal{G}$ be a series of neural operations (convolutions, nonlinearities, normalization, etc), a residual block takes an input and adds it to its transformed version by $\mathcal{G}$, i.e., $h^{k}= h^{k} + \mathcal{G}(h^{k})$. The ease of training and the simple mathematical intuitions behind residual blocks made them ubiquitous in recent ConvNet architectures and led to successful application in several image and signal processing applications \cite{goodfellow2016deep}.  

\subsection{ConvNets for filtering IGWs}
Here, we aim to design a ConvNet, denoted by $\mathcal{F}$, to reconstruct a tide-free SSH field, referred to as $S_f$ from some $Input$, which may be given as a single or a series of SSH snapshots, possibly complemented by other observed sea surface fields, such as SST fields. Formally, the considering filtering issue comes to train $\mathcal{F}$ such that $S_f = \mathcal{F}(Input)$.


We consider the following architecture for ConvNet $\mathcal{F}$.
It consists of a first Conv layer followed by $N_{r}$ residual units as introduced in \cite{he2016identity} then a final regression Conv layer.  In this paper, we consider three choices for the input: i) \textbf{SSH}: Our ConvNet takes a single SSH snapshot as input, and outputs its filtered version. ii) \textbf{3SSH}: SWOT fast-sampling phase is expected to deliver SSH measurements on a daily cycle, we simulate this behaviour by considering as input SSH at time $t$ concatenated with SSH at $t-24h$ and $t+24h$. iii) \textbf{3SSH-SST}: we complement the three successive daily SSH snapshots as above with a SST field. The later is motivated by known relationships between SSH and SST features for some dynamical modes \cite{tandeo2013segmentation}.

To train our ConvNet, we consider a loss function, which 
combines a mean square error on the SSH and a mean absolute error on the Laplacian of the SSH:
\begin{equation}
\mathcal{L}(S_{f},\hat{S_{f}}) = ||S_{f}-\hat{S_{f}}||^2 + \alpha |\nabla^2 S_{f}-\nabla^2\hat{S_{f}}|,
\label{eq:loss}
\end{equation}
The rationale behind this choice is that SSH fields are rather smooth while vorticity fields contain a considerable amount of high frequency information that need to be preserved. In the machine learning community, it is known that the use of the MSE loss function leads to blurry images and that adding losses on image gradients or Laplacians can help sharpen the predictions \cite{mathieu2015deep}.
\section{Numerical Experiments}
\subsection{Experimental setup}

Weights of the convolutional layers were initialized with the Kaiming initialization \cite{he2015delving}, biases to zero. To ease the optimization of the ResNets at early stages, we found it useful to use the "Zero $\gamma$" heuristic \cite{he2019bag}. Slope of the LeakyReLUs is 0.2. The networks were coded using PyTorch, and trained using the ADAM optimizer with an initial learning rate of 3e-4. The learning rate is divided by 10 every 100 epochs. Early stopping is also used leading all the networks to converge after around 400 epochs. Experiments were run using a Nvidia Tesla V100 GPU with a batch size of 32. The hyperparameters $N_r$ and $\alpha$ were tuned using a grid search leading to $N_r=10$ and $\alpha=10^3$.

\subsection{Results}
We compare the performance of the variants of the proposed ConvNet. As baseline, we consider a linear $5\times5$ filtering of the SSH. The parameters of the linear filter are optimized with regard to loss function (\ref{eq:loss}) to provide a fair comparison with the proposed schemes. This linear filtering approach is referred to as \textbf{Baseline}.

\begin{figure}[t]
    \centering
    \includegraphics[scale=0.24]{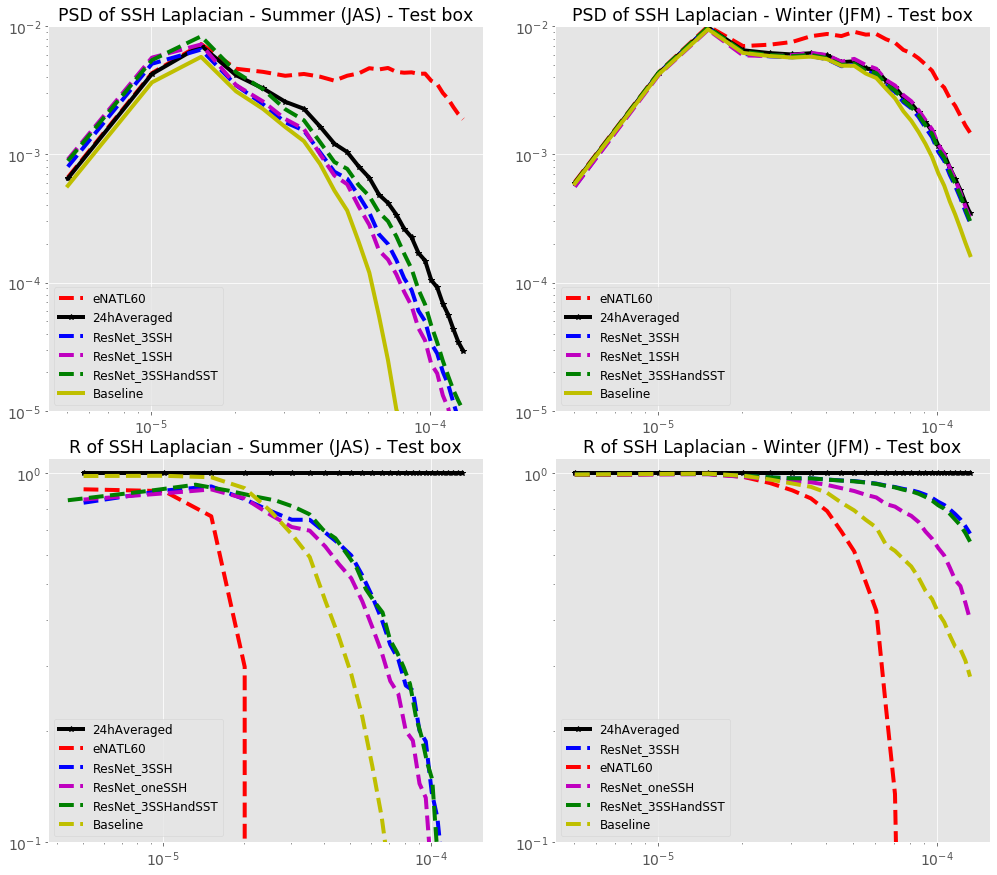}
    \caption{(Top) Power spectral density calculated on the Laplacian of SSH. (Bottom) Ratio $R$. Results on the JAS dataset are on the left, those on JFM on the right.}
    \label{fig:psd}
\end{figure}

Top panels of Figure \ref{fig:psd} show the power spectral density (PSD) functions of $\nabla^2$SSH resulting from the 4 competing methods. At large scale, the different approaches are close to the PSD of the reference data and differences start to appear at finer scales. For Summer data (left panel), the three proposed methods beat the baseline by a large margin, and unsurprisingly 3SSH and 3SSH-SST outperform 1SSH. For Winter data (right panel), the intensification of sea surface dynamics enhances non-wave processes at scales similar to IGWs, which makes the separation much more challenging. Still, the PSD shows that the three ConvNets outperform the baseline.

We further analyse the reconstruction performance according to 
ratio $R = 1 - \frac{E[(S_{ref}-S_{f})^2]}{E[S_{f}^2]}$ that reflects the relative quality of the estimation as a function of the spatial frequency.
$R$ values close to 1 indicate a perfect estimation. The $R$ values calculated on $\nabla^2$SSH are shown in the bottom panels of \ref{fig:psd}. For both JAS and JFM datasets we now clearly observe the differences between the introduced models. The baseline does not behave badly for a rough approximation, but is quickly outperformed by 1SSH and 3SSH. The addition of SST introduces a slight improvement over 3SSH. Examples of SSH filtered maps and their Laplacians can be found in this following Github repository \url{https://github.com/CIA-Oceanix/DetideNet} along with the codes used for this work. 


{}

\section{Conclusion}
In this work, we present a new deep learning based method to filter internal gravity waves from SSH data. Aiming for a proof of concept, this paper considered a numerical simulation as a testbed and has shown the relevance of the use of domain knowledge in the derivation of the loss function used for training the ConvNets. Experiments performed on winter and summer datasets prove that our method can yield competitive results with regard to simple linear filters. Results as presented are promising and call for more thorough investigation of the effect of adding other sources of noise that contaminate satellite altimeters, especially in the context of SWOT. This is the main subject of future work.

\section{acknowledgments}
This research was funded through CNES OST/ST and the SWOT Science Team. S. Metref is funded by ANR through contract number ANR-17- CE01-0009-01.
R. Fablet is supported by CNES (grant OSTST-MANATEE) and ANR (AI Chair OceaniX, EUR Isblue and Melody project).


\small
\bibliographystyle{IEEEbib}
\bibliography{IGWNET}

\end{document}